\documentclass[letterpaper, 10 pt, conference]{ieeeconf}  

\IEEEoverridecommandlockouts                              
\usepackage{graphicx} 
\usepackage{tikz}
\usetikzlibrary{shapes.geometric}
\usetikzlibrary{arrows}
\usepackage{forest}
\usetikzlibrary {shapes.multipart}
\usepackage{amsmath}
\usepackage{amssymb}
\usepackage{physics}
\usepackage{tcolorbox}

\usepackage{nicematrix}
\newcommand{\sgn}{\operatorname{sgn}}
\newcommand{\diag}{\operatorname{diag}}

\usepackage{hyperref}
\tikzstyle{sum} = [inner sep=1.5pt, draw, fill=white, circle, node distance=1cm]
\tikzstyle{block} = [draw, fill=white, rectangle, 
    minimum height=1.6em, minimum width=2.2em]
\tikzstyle{bluesum} = [inner sep=1.5pt, draw, fill=blue!10, circle, node distance=1cm]
\tikzstyle{blueblock} = [draw, fill=blue!10, rectangle, minimum height=1.6em, minimum width=2.2em]

\title{Determination of Bandwidth of Q-filter in Disturbance Observers \\ to Guarantee Transient and Steady State Performance \\ under Measurement Noise}

\author{Gaeun Kim${}^{*}$ and Hyungbo Shim${}$
\thanks{
  *Corresponding author
 }
\thanks{
  This work was supported by the National Research Foundation of Korea(NRF) grant funded by the Korea government(MSIT) (No. RS-2022-00165417).
 }
}

\begin{document}

\maketitle
\thispagestyle{empty}
\pagestyle{empty}

\begin{abstract}
    Q-filter-based disturbance observer (DOB) is one of the most widely used robust controller due to its design simplicity. Such simplicity arises from that reducing $\tau$, which is the time constant of low pass filters, 
not only ensures robust stability but also enhances \textit{nominal performance recovery}---ability to recover the trajectory of nominal closed-loop system.
However, in contrast to noise-free environment, excessively small $\tau$ can rather damage the nominal performance recovery under measurement noise. That is, minimizing $\tau$ is no longer immediately guaranteeing nominal performance recovery. Motivated by this observation, this paper concentrates on determination of $\tau$ to ensure transient and steady state performance.
This analysis uses Lyapunov method based on the coordinate change inspired by the singular perturbation theory.
As a result, we present an affordable noise level and open interval for $\tau$ that guarantees both the required performances. The analysis can also lead to theoretical demonstration on that excessively reducing $\tau$ is assured to achieve target performance only for noise-free case.    
\end{abstract}


\section{Introduction}

As a robust control against disturbances and parametric uncertainties, Q-filter-based disturbance observer (DOB) \cite{sariyildiz2019disturbance} has been used, attributing its straightforward design. One of the most important parameters in DOB is $\tau$,
which adjusts
 the cut-off frequency of low pass filters, called as \textit{Q-filter}. The reason of its importance stems from that a sufficiently small $\tau$ leads the overall system trajectory to follow that of the nominal closed-loop system arbitrarily close. It is demonstrated in noise-free by singular perturbation theory in prior studies \cite{Shim07CDCState,back2009inner}, and such capability of DOB is called as \textit{nominal performance recovery}. That is to say, it provides consistent design guideline that reducing $\tau$ yields better performance.

However, measurement noise is inevitable in more realistic situations and able to degrade the control performance.
Choosing $\tau$ excessively small can rather amplify the noise effect on internal signals and bother achieving control objective unlike noise-free case. In other words, simple and powerful design guideline based on minimizing $\tau$ is no longer applicable under measurement noise.

To tackle the noise sensitivity, previous researches proposed the modified structure of DOB for noise reduction \cite{xie2010high,jo2016noise}. There were remarkable results on noise attenuation, but they only dealt with high-frequency measurement noise separated with control bandwidth. Also, a few advantages of DOB were vanished, for instance, although \cite{jo2016noise} presented notable results under large noise amplitude, design procedure was involved in numerous parameters to be determined, and additional assumption on stability of nominal plant was required.

In this paper, in order to cope with the noise sensitivity of DOB, we concentrate on \textit{how to design $\tau$ in pre-configured DOB for guaranteeing desired performance under measurement noise}. This approach offers compatibility, as existing DOB can be utilized, and flexibility, as engineers can simply adjust the single parameter $\tau$ depending on control context. While this idea was employed in the framework of high-gain observer in \cite{ahrens2009high,prasov2012nonlinear}, it has not yet been explored in the context of DOB at the best of our knowledge.

This analysis is conducted by Lyapunov method based on the coordinate change inspired by the singular perturbation theory. Through the analysis, we present an affordable noise level and open interval for $\tau$ that guarantees both the desired transient and steady state nominal performance recovery. Furthermore, it is theoretically shown that extreme reduction of $\tau$ to achieve arbitrary target performance is effective only in the absence of noise as a result of the analysis.

The remainder of this paper is organized as follows: Section 2 describes the system and problem formulation. Section 3 reviews nominal performance recovery in noise-free case, and general case is handled in Section 4 as the main contribution. Section 5 concludes the paper.

\textit{Notation:} The set of real, nonnegative real, and positive real numbers are denoted by $\mathbb R$, ${\mathbb R}_{\geq0}$, and ${\mathbb R}_{>0}$, respectively. For a sequence $v_1,...,v_n$ of vectors or scalars, we define $[v_1;\cdots ;v_n]:=[v_1^\top \cdots v_n^\top]^\top$. A zero vector and a zero matrix is $0_i \in {\mathbb R}^i$ and $0_{i\times j}\in {\mathbb R}^{i\times j}$, respectively. A vector whose entries are all zero, except $i$-th entry that equals 1 is $\hat{\mathbf e}_i$ of appropriate size. The identity matrix of size $m \times m$ is $I_m$. For a vector $x$ and a matrix $A$, $\|x\|$ and $\| A \|$ denote the Euclidean norm and the induced matrix 2-norm, respectively. Any positive definite matrix $P$ is assumed to be symmetric and denoted as $P \succ 0$. For any $C \in {\mathbb R}^{1 \times i}$, matrices $A_i$, $B_i$, and $M_i^j(C)$ are
\begin{align*}
    A_i & := \begin{bmatrix}
    0_{i-1} & I_{i-1} \\
    0   &  0_{i-1}^\top
\end{bmatrix} \in {\mathbb R}^{i \times i},\, 
B_i := \begin{bmatrix}
    0_{i-1} \\
    1
\end{bmatrix} \in {\mathbb R}^{i \times 1},\\
M_i^j(C) & :=\left[C ; C A_i ; \cdots ;  C A_i^{j-1}\right] \in {\mathbb R}^{j \times i}.
\end{align*}


\section{Problem Formulation} \label{sec:2}

\subsection{Given system formulation} \label{sec:2:1}

Consider a single-input single-output (SISO) uncertain system $\mathrm P$ depicted in Fig. 1 described as
\begin{align}
\dot{x} & = A_\nu x +B_\nu \left(\phi^\top x + \psi^\top z +f_{\mathrm d}(x,z)+ g\cdot(u+d) \right) ,  \nonumber\\
\dot{z} & = Sz + G x_1 ,   \label{Real plant} \\
y & = \hat{\mathbf e}_1^\top x , \nonumber
\end{align} where $x\in {\mathbb R}^\nu$ and $z\in {\mathbb R}^{n-\nu}$ are the states, and $u \in {\mathbb R}$ is the control input. The output is $y\in {\mathbb R}$, but the measured output is mixed with measurement noise $v \in {\mathbb R}$ in Fig. 1. The disturbance $d\in {\mathbb R}$ is of $C^1$, and $d$ and $\dot d$ are uniformly bounded. System parameters $\phi$, $\psi$, $S$, and $G$ of appropriate dimension are unknown, but $S$ is assumed to be Hurwitz. The relative degree of the system (\ref{Real plant}) is $\nu>0$, so $g\in [g^-, g^+]$ is unknown nonzero scalar where lower bound $g^-$ and upper bound $g^+$ are known, and both $g^-$ and $g^+$ have the same sign. 
A function $f_{\mathrm d}$ represents unknown state dependent disturbance but assumed to be differentiable with respect to $t$ and locally Lipschitz with respect to $x$ and $z$.

The nominal model of $\mathrm P$, called as $\bar{\mathrm P}$, is written as
\begin{subequations} \label{nominal plant}
\begin{align}
\dot{x}_{\mathrm n} & = A_\nu x_{\mathrm n} +B_\nu \left(\bar{\phi}^\top x_{\mathrm n} + \bar{\psi}^\top \bar{z}_{\mathrm n} + \bar{g}u_{\mathrm n} \right)  \label{nominal1} \\
\dot{\bar z}_{\mathrm n} & = \bar S \bar z_{\mathrm n} + \bar G y_{\mathrm n} \label{nominal2}\\
y_{\mathrm n} & = \hat{\mathbf e}_1^\top x_{\mathrm n} \label{nominal-out}
\end{align}
\end{subequations} where $x_{\mathrm n}\in {\mathbb R}^\nu$ and $\bar z_{\mathrm n}\in {\mathbb R}^{n-\nu}$ are the states of $\bar{\mathrm P}$. The input and output are $u_{\mathrm n} \in {\mathbb R}$ and $y_{\mathrm n} \in {\mathbb R}$, respectively. System parameters $\bar \phi$, $\bar \psi$, $\bar g$, $\bar S$, and $\bar G$ are counterparts of $ \phi$, $\psi$, $g$, $ S$, and $G$, respectively. Initial conditions $x_{\mathrm n}(t_0)$ and $z_{\mathrm n}(t_0)$ are the same as $x(t_0)$ and $z(t_0)$.

The linear outer-loop output-feedback controller $\mathrm C$ has designed with respect to the nominal model $\bar{\mathrm P}$ as follows
\begin{subequations} \label{outer-loop controller-nominal}
\begin{align}
\dot{\theta}_{\mathrm n} & = J\theta_{\mathrm n} + K(r - y_{\mathrm n}) && \in {\mathbb R}^{n_{\mathrm c}}, \label{out1} \\
u_{\mathrm n} & = L\theta_{\mathrm n}+D(r - y_{\mathrm n})  && \in {\mathbb R}, \label{out-out}
\end{align}
\end{subequations} where $\theta_{\mathrm n}$ is the $n_{\mathrm c}$-dimensional controller state for $n_{\mathrm c}\geq 0$, and $u_{\mathrm n}$ is the output. The matrices $J$, $K$, $L$, and $D$ of appropriate dimension are constant. The reference input $r \in {\mathbb R}$ is of $C^1$, and $r$ and $\dot r$ are uniformly bounded. Assume that (\ref{outer-loop controller-nominal}) is well-designed so that the nominal closed-loop system (\ref{nominal plant})--(\ref{outer-loop controller-nominal}) is asymptotically stable.

Since the outer-loop controller (\ref{outer-loop controller-nominal}) is in the real closed-loop system as in Fig. 1, actual $\mathrm C$ takes the form
\begin{subequations} \label{Outer-loop controller}
\begin{align}
\dot{\theta} & = J\theta + K(r - y-v) && \in {\mathbb R}^{n_{\mathrm c}}, \\
u_{\mathrm r} & = L\theta+D(r - y-v) && \in {\mathbb R},
\end{align}
\end{subequations} where $\theta$ and $u_{\mathrm r}$ are the controller state and output, respectively, and $\theta_{\mathrm n}(t_0)$ of the controller in the nominal closed-loop system (\ref{nominal plant})--(\ref{outer-loop controller-nominal}) equals to $\theta(t_0)$ in (\ref{Outer-loop controller}).

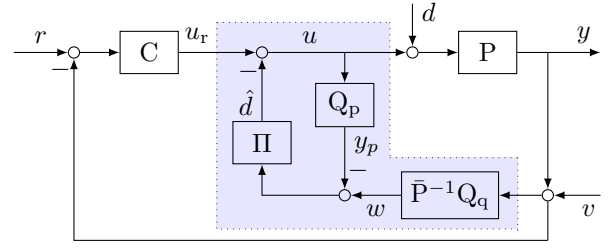
\begin{figure}
    \centering

    \begin{tikzpicture}[auto, node distance=0.5cm, >=latex]
        
        \fill[blue!10] (2.7,0.4) rectangle (5,-2.35);
        \fill[blue!10] (5,-2.35) rectangle (6.7,-1.4);
        
        \draw[dotted, line width=0.1mm] (2.7,0.4) -- (5,0.4);
        \draw[dotted, line width=0.1mm] (5,0.4) -- (5,-1.4);
        \draw[dotted, line width=0.1mm] (5,-1.4) -- (6.7,-1.4);
        \draw[dotted, line width=0.1mm] (6.7, -1.4) -- (6.7,-2.4);
        \draw[dotted, line width=0.1mm] (6.7,-2.35) -- (2.7,-2.35);
        \draw[dotted, line width=0.1mm] (2.7,0.4) -- (2.7,-2.35);

        \node[coordinate, name = referenceInput]{};
        \node[sum, right of = referenceInput, node distance = 0.8cm, name = sumOuter]{};
        \node[block, right of = sumOuter, node distance = 1cm, name = controllerOuter]{$\mathrm C$};
        \node[bluesum, right of = controllerOuter, node distance = 1.5cm, name = sumInner]{}; 
        \node[sum, right of = sumInner, node distance = 2cm, name = sumDist]{};
        \node[coordinate, right of = sumInner, node distance = 1.1cm, name = dobJunc]{};
        \node[coordinate, above of = sumDist, node distance = 0.65cm, name = disturbance]{};
        \node[block, right of = sumDist, node distance = 1cm, name = plant]{$\mathrm P$};
        \node[coordinate, right of = plant, node distance = 1.5cm, name = output]{};
        \node[coordinate, right of = plant, node distance = 0.8cm, name = outputJunc]{};
        \node[sum, below of = outputJunc, node distance = 1.9cm, name = noiseJunc]{}{};
        \node[coordinate, below of = output, node distance = 1.9cm, name = noise]{}{};
        \node[blueblock, left of = noiseJunc, node distance = 1.3cm, name = QA]{$\bar{\mathrm P}^{-1} {\mathrm Q_{\mathrm q}}$}; 
        \node[blueblock, below of = dobJunc, node distance = 0.7cm, name = QB]{${\mathrm Q}_ {\mathrm p}$};
        \node[blueblock, below of = sumInner, node distance = 1.2cm, name = sat]{$\Pi$};
        \node[bluesum, below of = QB, node distance = 1.2cm, name = sumQ]{};
        \node[coordinate, below of = outputJunc, node distance = 2.5cm, name = temp]{};
        
        \draw[->] (referenceInput) -- (sumOuter){} node[pos = 0.5]{$r$};
        \draw[->] (sumOuter) -- (controllerOuter){} node[pos = 0.5]{};
        \draw[->] (controllerOuter) -- (sumInner) node[pos = 0.24] {$u_{\mathrm r}$};
        \draw[->] (sumInner) -- (sumDist) node[pos = 0.3] {$u$};
        \draw[->] (sumDist) -- (plant) {};
        \draw[->] (plant) -- (output) node[pos=0.8]{$y$};
        \draw[->] (dobJunc) -- (QB){};
        \draw[->] (QB) -- (sumQ) node[inner sep=1pt, pos = 0.7]{$-$} node[pos=0.3]{$y_p$};
        \draw[->] (QA) -- (sumQ) node[pos = 0.5] {$w$};
        \draw[->] (sumQ) -| (sat) node [inner sep=1pt, pos = 0.94]{} node[pos = 0.75]{};
        \draw[->] (sat) -- (sumInner) node [inner sep=1pt, pos = 0.8]{$-$} node[pos = 0.3]{$\hat{d}$};
        \draw[->] (disturbance) -- (sumDist) node[pos = 0.2]{$d$} node[pos = 0.99] {};
            \draw[->] (noise) -- (noiseJunc) node[pos = 0.25]{$v$} node[inner sep=1pt, pos = 0.85]{} ;
        \draw[->] (outputJunc) -- (noiseJunc);
        \draw[->] (noiseJunc) -- (QA) node[pos = 0.25]{};
        \draw[-] (noiseJunc) -- (temp);
        \draw[->] (temp) -| (sumOuter) node[inner sep=1pt, pos = 0.97]{$-$};
    \end{tikzpicture}
    \caption{Closed-loop system with DOB (blue block)} 
    \label{fig:l}
\end{figure} 

Due to the disturbances and uncertainties in $\mathrm P$, it is clear that $y$ is different from the nominal closed-loop output $y_{\mathrm n}$. To enforce $y$ to get closer to $y_{\mathrm n}$, the inner-loop controller DOB is introduced as shown in Fig. 1. Control objective of DOB is to generate $\hat{d} \in {\mathbb R}$ in Fig. 1 so that $[x;\theta]$ in the real plant (\ref{Real plant}) and the actual outer-loop controller (\ref{Outer-loop controller}) is arbitrarily close to $[x_{\mathrm n};\theta_{\mathrm n}]$.

Let $\chi_{\mathrm n} := [x_{\mathrm n};\bar z_{\mathrm n}; \theta_{\mathrm n}; z_{\mathrm n}]\in {\mathbb R}^{n_{\mathrm e}}$ consist of the nominal closed-loop system (\ref{nominal plant})--(\ref{outer-loop controller-nominal}), and the virtual $z_{\mathrm n}$:
\begin{equation} \label{alone-zero}
 \dot{z}_{\mathrm n} = Sz_{\mathrm n} + G y_{\mathrm n} ~ \in {\mathbb R}^{n-\nu},~~~~z_{\mathrm n}(t_0) = z(t_0).
\end{equation} Then, virtual $\chi_{\mathrm n}$-dynamics can be expressed as follows 
\begin{equation} \label{chi-N}
    \dot \chi_{\mathrm n} = {\mathbf A}_{\mathrm s}\chi_{\mathrm n} + {\mathbf B}r ~~~\in {\mathbb R}^{n_{\mathrm e}},
\end{equation} where ${\mathbf A}_{\mathrm s}\in {\mathbb R}^{n_{\mathrm e} \times n_{\mathrm e}}$ and ${\mathbf B}\in {\mathbb R}^{n_{\mathrm e}}$ are proper matrices that make (\ref{chi-N}) identical to the nominal closed-loop system augmented by the virtual internal dynamics (\ref{nominal plant}), (\ref{outer-loop controller-nominal}), and (\ref{alone-zero}). Note that ${\mathbf A}_{\mathrm s}$ is Hurwitz, relying on asymptotic stability of the nominal closed-loop system and Hurwitzness of $S$.

\subsection{Configuration of DOB in state space}

Now, we introduce configuration of DOB in the blue block of Fig. 1, consisting of three components: (i) a combination of the inverse nominal model and Q-filter $\bar{\mathrm P}^{-1} {\mathrm Q_{\mathrm q}}$ which receives $y+v$; (ii) another Q-filter ${\mathrm Q_{\mathrm p}}$ receiving $u$; and (iii) a saturation function $\Pi$\footnote{Saturation function is defined as $\Pi({\mathsf x}):= \begin{cases}
    \sgn({\mathsf x}) \bar s,& \text{if } |{\mathsf x}| > \bar s\\
    {\mathsf x},              & \text{if } |{\mathsf x}| \leq \bar s \\
\end{cases}$ where $\bar s>0$ is saturation level. The saturation level $\bar s$ is required to be set up sufficiently large.}. Before representing DOB in state space, we share that the considered Q-filters are the general form in transfer function: \begin{equation*}
    {\mathrm Q}_{\mathrm q}(s) = {\mathrm Q}_{\mathrm p}(s) = \frac{c_{l-m}(\tau s)^{l-m}+\cdots +c_1(\tau s)+a_0}{(\tau s)^l + a_{l-1}(\tau s)^{l-1}+\cdots+ a_0}.
\end{equation*}

First, the combined block $\bar{\mathrm P}^{-1} {\mathrm Q_{\mathrm q}}$ is expressed as
\begin{subequations} \label{QQ}
\begin{alignat} {4}
    \dot{\bar{z}} & = \bar{S}\bar{z}+\bar{G}(y+v)  && \in {\mathbb R}^{n-\nu}, && \label{ss_z} \\
    \dot{q} & = A_lq - \frac{a_0}{\tau^l}T_\tau \hat{\mathbf e}_l (q_1-y-v) && \in {\mathbb R}^l, \label{ss-q}\\
    w & = -\frac{1}{\bar{g}}\left( \bar{\psi}^\top \bar{z}+ \bar{\phi}^\top M_l^\nu(C_\tau)T_\tau ^{-1}q \right) && +\frac{1}{\bar{g}} C_\tau A_l^\nu T_\tau ^{-1} q ,\label{w}
\end{alignat}\end{subequations} \begin{equation*}
    \text{where } C_\tau = \frac{1}{a_0} \left[a_0 \; c_1\tau \;\cdots \; c_{l-m}\tau^{l-m} ~\; 0^\top_{m-1} \right] \in {\mathbb R}^{1 \times l},\end{equation*}
\begin{equation*}
\text{and }T_\tau = \frac{1}{a_0} \begin{bmatrix}
a_0 & a_1 \tau  & \cdots & a_{l-1} \tau^{l-1} \\
   0    &    \ddots & \ddots & \vdots \\ 
    \vdots   &    \ddots    & a_0 & a_1 \tau \\
    0   &  \cdots   &    0    & a_0 \\
\end{bmatrix} \in {\mathbb R}^{l \times l}, 
\end{equation*} for $l\geq m >\nu$, some coefficients $\{ a_0,..., a_{l-1}\}\subset {\mathbb R}_{>0}$, and $\{ c_1,..., c_{l-m}\} \subset {\mathbb R}_{\geq 0}$. The key parameter $\tau>0$ will be determined elaborately. 
The DOB state $\bar z$ represents the internal dynamics of $\bar{\mathrm P}$ associated with $y+v$. The block output $w\in {\mathbb R}$ consists of $\bar z$ and Q-filter state $q$.

Another Q-filter ${\mathrm Q_{\mathrm p}}$ is expressed in the controllable canonical form: \begin{subequations} \label{QP}
\begin{alignat}{3}
\dot{p} & = A_lp+\hat{\mathbf e}_l \left( - \frac{a_0}{\tau^l} \hat{\mathbf e}_1^\top T_\tau p + u \right) && \in {\mathbb R}^l, \label{ss-p}\\
    y_p & = \frac{a_0}{\tau^l}C_\tau p && \in {\mathbb R}, \label{yp}
\end{alignat}
\end{subequations} where $p$ is the Q-filter state, and $y_p$ is the Q-filter output.

\textit{Remark 1:} Note that $m> \nu$ implies that the relative degree of Q-filter is strictly larger than that of the nominal plant $\bar {\mathrm P}$, i.e., $\bar {\mathrm P} ^{-1}{\mathrm Q}_{\mathrm q}$ in frequency domain is strictly proper. The reason of excluding $m=\nu$ lies on that it causes inherent noise amplification into the output $w$. When Q-filter is designed as $m=\nu$, $c_{l-m}\tau^{-m} \hat {\mathbf e}_l (y+v)$ is added in $\bar {\mathrm P} ^{-1}{\mathrm Q}_{\mathrm q}$ output (\ref{w}) because of $C_\tau A_l^{\nu-1} \hat {\mathbf e}_l \neq 0$. It makes internal signals of DOB directly exposed to noise effect and get worse as $\tau\rightarrow 0$. To avoid such extreme occasion, we simply brings Q-filter of $m>\nu$.

\subsection{Performance metric and Problem statement}

In this section, problem that we focus on is formally stated. To begin with, a metric for quantitatively evaluating the performance is defined as follows
\begin{equation} \label{errdef}
    e := \chi-\chi_{\mathrm n}~~~~~~\in {\mathbb R}^{n_{\mathrm e}},
\end{equation} where $\chi\in {\mathbb R}^{n_{\mathrm e}}$ is $\chi 
= [x;\bar z;\theta;z]$ as the counterpart of the virtual system state $\chi_{\mathrm n}$ in (\ref{chi-N}). Components of $\chi$ are from the real plant (\ref{Real plant}), outer-loop controller (\ref{Outer-loop controller}), and inverse nominal model inside DOB (\ref{ss_z}). It is evident that $e=0$ implies complete recovery of nominal performance.
Moreover, $e$-dynamics can be expressed as follows by defining new variable ${\mathbf d}$: \begin{equation} \label{sec2:3-err:dyna}
    \dot{e} = {\mathbf A}_{\mathrm s}e +g\hat{\mathbf e}_\nu \left( {\mathbf d}- \hat{d}\right) - Nv,
\end{equation} \begin{align*} \text{where}~~&
    {\mathbf d}
    =\frac{1}{g}\left((\phi^\top -\bar \phi^\top) x + \psi^\top z-\bar \psi^\top \bar z \right) \\ & \qquad \qquad \qquad + \frac{1}{g} \left( f_{\mathrm d}(x,z)+gd  +(g-\bar{g})u_{\mathrm r} \right),
\end{align*} and $N=[\bar{g}DB_\nu; - \bar G;K; 0_{n-\nu}] \in {\mathbb R}^{n_{\mathrm e}}$.

Meanwhile, an assumption on the boundedness of the measurement noise $v$ is introduced.

\textit{Assumption 1:} The measurement noise $v$ is bounded by $\mu$, i.e., $|v(t)| \leq \mu,~\forall t\geq t_0$, and may not be continuous nor differentiable.

In this paper, we address the following performance guarantee problem.

\textit{Problem 1: } For any desired $\varepsilon_{\mathrm U}>0$ and $\varepsilon_{\mathrm T} > \varepsilon_{\mathrm U}$, suggest affordable bound of the measurement noise, and design DOB parameter $\tau$ in Q-filters (\ref{QQ}) and (\ref{QP}) such that the performance metric $e$ satisfies: \begin{subequations}
\begin{align} 
     \text{\textit{(Transient)} }~~ &\|e(t)\| <\varepsilon_{\mathrm T},~~~\forall t \geq t_0, \label{prob1} \\
    \text{\textit{(Steady state)} }~~ & \lim_{t\rightarrow \infty}\|e(t)\| <\varepsilon_{\mathrm U}. \label{prob2}
\end{align}
\end{subequations}

The explicit answers to Problem 1 are in Proposition 1 for ideal noise-free case and Theorem 1 for general case, affected by the measurement noise.

\section{Review of Performance Recovery for Noise-free Case}

Before dealing with the measurement noise, we first solve Problem 1 only for noise-free case. In fact, the DOB analysis using the singular perturbation theory \cite{kokotovic1999singular} can directly resolve Problem 1, as in \cite{back2009inner}. Key point to apply the singular perturbation theory is finding a suitable state representation written in the standard singular perturbation form. Thus, noting that $v$ is identically zero only in this section, we present the lemma on coordinate change to transform the overall closed-loop system into the standard singular perturbation form

\textit{Lemma 1:} The noise-free overall closed-loop system with the states $q$, $p$, and $e$ in (\ref{QQ})--(\ref{errdef}) is transformed into the singularly perturbed system by the coordinate change as follows 
\begin{subequations} \label{coordi}
    \begin{align} 
    \xi &:= \tau^{-\nu-1}\Delta\left(q-[x;~0_{l-\nu}] \right)-\hat{\mathbf e}_{\nu+1} \hat{\mathbf e}_\nu^\top\Xi^\ast_\xi, \label{coordi1} \\
    \zeta & := \tau^{-l-1} \Delta p -\hat{\mathbf e}_1\Xi^\ast_\zeta, \label{coordi2}
\end{align} \end{subequations} where $\Delta=\diag(\tau,...,\tau^l)$, $\Xi^\ast_\zeta = (u_{\mathrm r}- {\mathbf d} )/a_0\in{\mathbb R}$, and $\Xi^\ast_\xi ={\mathbf A}_{\mathrm s} e-({\mathbf A}_{\mathrm s} \chi_{\mathrm n} + {\mathbf B}r) \in {\mathbb R}^{n_{\mathrm e}}$, which implies $\hat{\mathbf e}_\nu^\top \Xi_\xi^\ast = \bar{\phi}^\top x + \bar{\psi}^\top \bar{z} + \bar{g}  u_{\mathrm r}$. The entire system written in the standard singular perturbation form is
\begin{align} 
    \tau \dot{\xi} & = {\mathbf A}_{\mathrm 11} \xi + {\mathbf A}_{\mathrm 12} \zeta  -\tau g\hat{\mathbf e}_\nu \left( {\mathbf C}_\tau \xi  + \bar{\mathbf C}_\tau \Xi_\xi^\ast \right) \nonumber \\ 
    & \quad \, - \tau \hat{\mathbf e}_{\nu+1} \hat{\mathbf e}_\nu^\top \dot \Xi^\ast_\xi,  \label{noise-free-singular}  \\
    \tau \dot{\zeta} & = {\mathbf A}_{\mathrm 21} \xi + {\mathbf A}_{\mathrm 22}\zeta +\tau \hat{\mathbf e}_l\left({\mathbf C}_\tau \xi  + \bar{\mathbf C}_\tau \Xi_\xi^\ast\right)- \tau \hat{\mathbf e}_1 \dot{\Xi}^\ast_\zeta, \nonumber \\
    \dot{e} & = {\mathbf A}_{\mathrm s}e + {\mathbf F}_{\mathrm 1}\xi + {\mathbf F}_{\mathrm 2}\zeta + \tau g\hat{\mathbf e}_\nu \left( {\mathbf C}_\tau \xi  + \bar{\mathbf C}_\tau \Xi_\xi^\ast \right), \nonumber
\end{align}

\begin{align*}
    \text{where } {\mathbf A}_{\mathrm 11} & = A_l - a_0T_1\hat{\mathbf e}_l\hat{\mathbf e}_1^\top +\frac{g}{\bar g}\hat{\mathbf e}_\nu C_1A_l^\nu T_1^{-1} , \\
    {\mathbf A}_{\mathrm 12} & = -ga_0\hat{\mathbf e}_\nu C_1 ,\\
    {\mathbf A}_{\mathrm 21} & = -\frac{1}{\bar g}\hat{\mathbf e}_l C_1A_l^\nu T_1^{-1}, \quad {\mathbf A}_{\mathrm 22} = A_l + a_0\hat{\mathbf e}_l \bar{C}_1, \\
    {\mathbf C}_\tau & = \frac{1}{\bar g} \bar \phi^\top \diag (\tau^{\nu-1},...,1) M_l^\nu(C_1)T_1 ^{-1}, \\
    \bar{\mathbf C}_\tau & = \frac{1}{\bar g}\bar \phi^\top M_l^\nu(\tau^{-1}\bar{C}_\tau)T_\tau^{-1}[\hat{\mathbf e}_2 \cdots  \hat{\mathbf e}_{\nu+1}  0_{l \times (n_{\mathrm e} -\nu)}],\\
    {\mathbf F}_{\mathrm 1} & = -\frac{g}{\bar g} \hat{\mathbf e}_\nu C_1 A_l^\nu T_1 ^{-1}, \quad {\mathbf F}_{\mathrm 2} = ga_0 \hat{\mathbf e}_\nu C_1,\\
    \bar C_\tau & =C_\tau-\hat{\mathbf e}_1^\top T_\tau,
\end{align*} and $C_1$, $T_1$, and $\bar C_1$ are $C_\tau$, $T_\tau$, and $\bar C_\tau$ with $\tau=1$, respectively.

\textit{Proof: } Lemma 1 is proved in the demonstration for noisy case of Section 4 by replacing $v\equiv0$. See Appendix A.  $\square$.

 We partition the singularly perturbed system (\ref{noise-free-singular}) into two subsystems: fast subsystem with $\xi$ and $\zeta$, whose dynamics become faster as $\tau \rightarrow 0$, and slow subsystem with $e$ otherwise. The following assumption on coefficients of Q-filter should be satisfied for exponential stability of fast subsystem. 

\textit{Assumption 2:} Q-filter coefficients $\{ a_0,..., a_{l-1}\}$ and $\{ c_1,..., c_{l-m}\}$ in the matrices ${\mathbf A}_{\mathrm 11}$, ${\mathbf A}_{\mathrm 12}$, ${\mathbf A}_{\mathrm 21}$, and ${\mathbf A}_{\mathrm 22}$ are designed so that below the block matrix is Hurwitz: \begin{align*} {\mathbf A}_{\mathrm f} :=
\begin{bmatrix}
        {\mathbf A}_{\mathrm 11} && {\mathbf A}_{\mathrm 12} \\
        {\mathbf A}_{\mathrm 21} && {\mathbf A}_{\mathrm 22}
    \end{bmatrix},
\end{align*} for every uncertain but bounded $g \in [g^-, g^+]$.

By Lemma 1 and Assumption 2, the solution of Problem 1 for noise-free case naturally arises based on Tikhonov's theorem \cite{hoppensteadt1966singular}. See \cite{Shim07CDCState,back2009inner} if you need more details on DOB analysis by the singular perturbation theory.

\textit{Proposition~1}~\textit{[3, ~Theorem~1]:} Let Assumption 2 be hold. Then, for any desired $\varepsilon_{\mathrm U}>0$ and $\varepsilon_{\mathrm T} > \varepsilon_{\mathrm U}$, there exists $\bar\tau >0$ such that for all $\tau \in (0, \bar \tau)$, the performance metric $e$ in the absence of the measurement noise in (\ref{noise-free-singular}) satisfies (\ref{prob1}) and (\ref{prob2}) in Problem 1.

\section{Nominal Performance Recovery with Noisy Measurement}

From now on, we consider the measurement noise again which satisfies Assumption 1 in Section 2.3. First, we observe the difficulty to bring the singular perturbation theory under measurement noise. After that, answer for Problem 1 and explicit design on $\mu^\ast$, $\underline{\tau}$, and $\bar \tau$ of Problem 1 are handled in Section 4.1 and 4.2, respectively.

To observe whether the singular perturbation theory is still valid, the same coordinate transformation (\ref{coordi}) is applied in the presence of measurement noise. Reusing the notations and most of the definitions in Lemma 1, we can obtain the overall system written with $\xi$ and $\zeta$:
\begin{subequations} \label{noisy-singular}
\begin{align}
    \tau \dot{\xi} & =  {\mathbf A}_{\mathrm 11} \xi + {\mathbf A}_{\mathrm 12} \zeta  -\tau g\hat{\mathbf e}_\nu \left( {\mathbf C}_\tau \xi  + \bar{\mathbf C}_\tau \Xi_\xi^\ast \right) \nonumber \\
    & \quad ~ - \tau \hat{\mathbf e}_{\nu+1} \hat{\mathbf e}_\nu^\top \dot \Xi^\ast_\xi + \frac{a_0}{\tau^\nu}T_1 \hat{\mathbf e}_l v  + {\mathbf N}_\xi v, \label{noisy-singular-xi}\\
    \tau \dot{\zeta} & = {\mathbf A}_{\mathrm 21} \xi + {\mathbf A}_{\mathrm 22}\zeta +\tau \hat{\mathbf e}_l\left({\mathbf C}_\tau \xi  + \bar{\mathbf C}_\tau \Xi_\xi^\ast\right) \nonumber \\ 
    & \quad ~ - \tau \hat{\mathbf e}_1 \dot{\Xi}^\ast_\zeta + {\mathbf N}_\zeta v, \label{noisy-singular-zeta} \\
    \dot{e} & = {\mathbf A}_{\mathrm s}e + {\mathbf F}_{\mathrm 1}\xi + {\mathbf F}_{\mathrm 2}\zeta + \tau g\hat{\mathbf e}_\nu \left( {\mathbf C}_\tau \xi  + \bar{\mathbf C}_\tau \Xi_\xi^\ast \right) \nonumber \\
    & \quad ~ + {\mathbf N}_{\mathrm e}v, \label{noisy-perfor}
\end{align} \end{subequations} where ${\mathbf N}_\xi =gD\hat{\mathrm e}_\nu \in {\mathbb R}^l$, ${\mathbf N}_\zeta =-D\hat{\mathrm e}_l \in {\mathbb R}^l$, and ${\mathbf N}_{\mathrm e} = (\bar g- g)D\hat{\mathrm e}_\nu-N \in {\mathbb R}^{n_{\mathrm e}}$. We redefine $\Xi^\ast_\zeta$ as $\Xi^\ast_\zeta = (u_{\mathrm r}- {\mathbf d} + (\bar g D v)/g)/a_0\in{\mathbb R}$, and $\Xi^\ast_\xi$ implies $\hat{\mathbf e}_\nu^\top \Xi_\xi^\ast = \bar{\phi}^\top x + \bar{\psi}^\top \bar{z} + \bar{g} ( u_{\mathrm r}+Dv)$. As a remark, both $\Xi^\ast_\xi$ and $\Xi^\ast_\zeta$ are differentiable, since the noise terms inside $u_r$ are eliminated as for the definition of $\mathbf d$ in (\ref{sec2:3-err:dyna}). How to obtain the transformed system (\ref{noisy-singular}) by executing the coordinate change (\ref{coordi}) can be found in Appendix A.

It is revealed that the transformed system (\ref{noisy-singular}) does not take the standard singular perturbation form due to the term including $\tau^{-\nu}$ in (\ref{noisy-singular-xi}), generated by the measurement noise. To be specific, the fast dynamics (\ref{noisy-singular-xi})--(\ref{noisy-singular-zeta}) become faster as $\tau \rightarrow 0$ with the exponential decay rate of order $\tau^{-1}$. However, the nonvanishing perturbation term with $\tau^{-\nu}$ in (\ref{noisy-singular-xi}) also becomes dominant and scales as $\tau^{-\nu-1}$ of higher than the decay rate. Therefore, excessively small $\tau$ rather causes performance degradation, however, still sufficiently small $\tau$ is required for weakening the other perturbations and time-scale separation.

\subsection{Performance guarantee by Lyapunov method}

Instead of the singular perturbation theory, we employ the Lyapunov analysis on each of the fast and slow subsystem and present the answer for Problem 1 under measurement noise.

\textit{Theorem 1:} Suppose that Assumption 1-2 hold. Then, for any desired $\varepsilon_{\mathrm U}>0$ and $\varepsilon_{\mathrm T} > \varepsilon_{\mathrm U}$, there is a threshold of noise level $\mu^\ast > 0$ such that for every $\mu<\mu^\ast$, there exist a lower bound $\underline{\tau} =\underline{\tau}(\mu) \geq 0$ and an upper bound $\bar \tau=\bar \tau(\mu)>\underline{\tau}$ such that for all $\tau \in (\underline{\tau}, \bar \tau)$, the performance metric $e$ under the measurement noise $v$ in (\ref{noisy-perfor}) satisfies transient (\ref{prob1}) and steady state (\ref{prob2}) performance in Problem 1.

\textit{Proof:} To begin with, combine the fast dynamics as $\eta:=[\xi; \zeta] \in {\mathbb R}^{2l}$ and express the overall system (\ref{noisy-singular}) with the stacked vector as
\begin{align*}
    \tau \dot \eta & = {\mathbf A}_{\mathrm f} \eta + \tau \check{\mathrm e} \check{\mathbf C}_\tau \eta + \tau \check{\mathrm e}  \bar{\mathbf C}_\tau \Xi^\ast _\xi - \tau \dot \Xi^\ast_\eta + \frac{1}{\tau^\nu}\pmb{\alpha}v  + {\mathbf N}_\eta v, \\
 \dot{e} & = {\mathbf A}_{\mathrm s}e + {\mathbf F}\eta  + \tau g\hat{\mathbf e}_\nu \check {\mathbf C}_\tau \eta +\tau g\hat{\mathbf e}_\nu \bar{\mathbf C}_\tau \Xi_\xi^\ast + {\mathbf N}_{\mathrm e}v, 
\end{align*} 
where $\check{\mathrm e}=- g\hat{\mathbf e}_\nu + \hat{\mathbf e}_{2l} \in {\mathbb R}^{2l}$, $\check{\mathbf C}_\tau = [{\mathbf C}_\tau ~0^\top_l]\in {\mathbb R}^{1 \times 2l}$, $\Xi^\ast_\eta=  \hat{\mathbf e}_{\nu+1} \hat{\mathbf e}_\nu^\top  \Xi^\ast_\xi + \hat{\mathbf e}_{l+1} \Xi^\ast_\zeta \in {\mathbb R}^{2l}$, $\pmb{\alpha}=a_0  [T_1\hat{\mathbf e}_l ;0_{l}] \in {\mathbb R}^{2l}$, ${\mathbf N}_\eta = [{\mathbf N}_\xi; {\mathbf N}_\zeta] \in {\mathbb R}^{2l}$, and ${\mathbf F} = [{\mathbf F}_1 ~ {\mathbf F}_2]\in {\mathbb R}^{n_{\mathrm e} \times 2l}$.

There exist $P_{\mathrm f}\succ 0$ and $P_{\mathrm s}\succ 0$ such that ${\mathbf A}_{\mathrm f}^\top P_{\mathrm f} +P_{\mathrm f}{\mathbf A}_{\mathrm f}=-I_{2l}$ and ${\mathbf A}_{\mathrm s}^\top P_{\mathrm s} +P_{\mathrm s}{\mathbf A}_{\mathrm s}=-I_\nu$ based on Assumption 2 and asymptotic stability of the virtual system (\ref{chi-N}), respectively. Maximal and minimal eigenvalues of $P_{\mathrm f}$ and $P_{\mathrm s}$ are denoted as $\lambda^{\mathrm f}_{\max}$, $\lambda^{\mathrm f}_{\min}$, $\lambda^{\mathrm s}_{\max}$, and $\lambda^{\mathrm s}_{\min}$, respectively.

Let $V_{\mathrm f}$ and $V_{\mathrm s}$ be defined as $V_{\mathrm f}=\eta^\top P_{\mathrm f}\eta$ and $V_{\mathrm s}=e^\top P_{\mathrm s} e$.
Take $c_{\mathrm U}>0$ and $c_{\mathrm T}>0$ such that 
$c_{\mathrm U}<\lambda^{\mathrm s}_{\min} \varepsilon_{\mathrm U}^2 < c_{\mathrm T}<\lambda^{\mathrm s}_{\min} \varepsilon_{\mathrm T}^2$, and let $\Omega_{\mathrm T}:= \{e:V_{\mathrm s}(e)\leq c_{\mathrm T} \}$ be compact due to the properness of $V_{\mathrm s}$. 

We are going to show Theorem 1 by guaranteeing i) $e \in \Omega_{\mathrm T}$ for all $t\geq t_0$, invoking that $e(t_0)=0$, and ii) $ \lim_{t \rightarrow \infty} V_{\mathrm s} < c_{\mathrm U}$.

Suppose $e\in \Omega_{\mathrm T}$, then there exist $\kappa_1,\,\kappa_2>0$ which are bounds of $\Xi_\xi^\ast$ and $\dot{\Xi}^\ast_\eta$, respectively. Boundedness of $\Xi_\xi^\ast$ is as for the boundedness of $\chi_{\mathrm n}$. Also, $\dot{\Xi}^\ast_\eta$ is bounded based on the boundedness of $\dot r$, $\dot{\chi}_{\mathrm n}$, $d$, $\dot{d}$, and $f_{\mathrm d}$, $\dot{f}_{\mathrm d}$ by local Lipshitzness and $e \in \Omega_{\mathrm T}$. Hence, $\dot V_{\mathrm f}$ is \begin{equation*}
    \begin{split}
        \dot{V}_{\mathrm f} 
  & \leq -\frac{V_{\mathrm f}}{\lambda_{\max}^{\mathrm f} \tau} +2 \sqrt{\check g^2 +1}\| \check{\mathbf C}_\tau\|V_{\mathrm f} \\
        &~~~~+\frac{1}{2 \lambda_{\max}^{\mathrm f}}\sqrt{V}_{\mathrm f} \left( \sigma_1 \|\bar{\mathbf C}_\tau \| +\sigma_2  +\frac{\sigma_3}{\tau^{\nu+1}}\mu+\frac{\sigma_4}{\tau}\mu  \right),
    \end{split}
\end{equation*}
where $\check g :=\max|g|$, $\sigma_1= 4 \sqrt{\check g^2 +1} \kappa_1 \lambda_{\max}^{\mathrm f} / (\lambda_{\min}^{\mathrm f})^{1/2}$, $\sigma_2 =4\kappa_2\lambda_{\max}^{\mathrm f} / (\lambda_{\min}^{\mathrm f})^{1/2}$, $\sigma_3= 4\|\pmb{\alpha}\|\lambda_{\max}^{\mathrm f} / (\lambda_{\min}^{\mathrm f})^{1/2}$, and $\sigma_4=4 \check g |D|\lambda_{\max}^{\mathrm f} / (\lambda_{\min}^{\mathrm f})^{1/2}$.

Let $\bar \tau_1$ be the smallest $\tau>0$ satisfying $\sigma_\tau:=1/(2\lambda_{\max}^{\mathrm f}\tau)- 2\sqrt{\check g^2 +1} \| \check{\mathbf C}_\tau\|=0$. Then for all $\tau<\bar \tau_1$, $\sigma_\tau>0$ holds, and there exists $\kappa_3>0$ such that $\|\bar{\mathbf C}_\tau \|\leq\kappa_3$, since $\check {\mathbf C}_\tau$ and $\bar{\mathbf C}_\tau$ comprise nonnegative integer powers of $\tau$. Thus, it can be shown that \begin{equation} \label{fast_invariant}
    \dot{V}_{\mathrm f} \leq -\sigma_\tau V_{\mathrm f},~~~~~~~~\text{for}~V_{\mathrm f} \geq \Sigma_{\mathrm f}(\mu,\tau)^2
\end{equation} where $\Sigma_{\mathrm f}:{\mathbb R}_{\geq0}\times{\mathbb R}_{>0}\rightarrow {\mathbb R}$ is defined as $\Sigma_{\mathrm f}(\mu, \tau)=\sigma_3  \tau^{-\nu} \mu+\sigma_4\mu+\sigma_5 \tau $ with $\sigma_5:=\sigma_1\kappa_3+\sigma_2$. Eqn. (\ref{fast_invariant}) implies that the set $\Sigma :=\{ \eta :V_{\mathrm f}(\eta)\leq \Sigma_{\mathrm f}(\mu,\tau)^2\}$ is positively invariant. Inside $\Sigma$, $\|\eta\|$ is bounded as $ \|\eta\| \leq \Sigma_{\mathrm f}(\mu,\tau) (\lambda_{\min}^{\mathrm f})^{-1/2}$. 
If $V_{\mathrm f}(\eta(t_0))\geq \Sigma_{\mathrm f}(\mu,\tau)^2$, then 
$\eta$ reaches the set within $[t_0, t_0+ {\mathsf t} (\tau)]$ where
\begin{equation} \label{sanserif-time}
     \frac{1}{\sigma_\tau}\ln{\frac{V_{\mathrm f}(\eta(t_0))}{(\sigma_5 \tau)^2}}\leq \frac{1}{\sigma_\tau}\ln{\frac{\lambda_{\max}^{\mathrm s}k_{\mathrm f}}{\sigma_5^2 \tau^{2(l+1)}}}=: {\mathsf t}(\tau),
\end{equation} where $k_{\mathrm f}\in{\mathbb R}$ satisfies $\|\tau^l \eta(t_0)\|^2 \leq k_{\mathrm f}$, regardless of $\tau$, as in (\ref{coordi}). If the set $\Omega_{\mathrm T}\times \Sigma$ is positively invariant, then the fast state $\eta$ should arrive in $\Sigma$ before $e$ escapes $\Omega_{\mathrm T}$ because $\eta(t_0) \in \Sigma$ is not assured. Due to the saturation function, there exists a finite time ${\mathsf t}^\ast$ independent of $\tau$, such that $e$ remains in $\Omega_{\mathrm T}$ for $[t_0, t_0+{\mathsf t}^\ast]$. Since ${\mathsf t}(\tau) \rightarrow 0$ as $\tau \rightarrow 0$ by $\sigma_\tau$, we take $\bar \tau_2>0$ so that ${\mathsf t}(\tau) < {\mathsf t}^\ast$ for every $\tau< \bar \tau_2$. Hence, i) $e \in \Omega_{\mathrm T}$ for all $t\geq t_0$ is shown if $\Omega_{\mathrm T}\times \Sigma$ is positively invariant.

Meanwhile, for $\eta \in \Sigma$ and $\tau<\min(\bar \tau_1, \bar \tau_2)$, there is $\kappa_4 = \|  {\mathbf F}\|+\check g \max_{\tau} \| {\mathbf C}_\tau\|$, and $\dot{V}_{\mathrm s}$ satisfies \begin{align}
        \dot{V}_{\mathrm s} & \leq -\frac{V_{\mathrm s}}{\lambda_{\max}^{\mathrm s}} +\frac{2 \lambda_{\max}^{\mathrm s} } {(\lambda_{\min}^{\mathrm s})^{1/2}} \sqrt{V}_{\mathrm s} 
        \left( \kappa_4 \Sigma_{\mathrm f} + \tau \check g \kappa_1 \kappa_3 + \kappa_5 \mu\right) \nonumber \\
        & =: -b_0 V_{\mathrm s} + \left(b_1 \Sigma_{\mathrm f}+ b_2 \tau+b_3 \mu \right)\sqrt{V_{\mathrm s}}, \label{Vs-dyna}
    \end{align} where  and $\kappa_5 = \check g |D| + | \bar g D| + \| N \|$. We restrict $e \in \partial \Omega_{\mathrm T}$, then $\dot{V}_{\mathrm s} < 0$ for $(e, \eta) \in \partial \Omega_{\mathrm T} \times \Sigma$ if the following is proved:
\begin{equation} \label{sigma bound} b_0\sqrt{c_{\mathrm T}}> b_1 \Sigma_{\mathrm f}+ b_2 \tau+b_3 \mu =: \bar \Sigma(\mu, \tau).
\end{equation}

Also for showing ii) $ \lim_{t \rightarrow \infty} V_{\mathrm s} < c_{\mathrm U}$, (\ref{Vs-dyna}) is modified to apply the comparison lemma \cite{khalil2002nonlinear} with $W(t)=\sqrt{V_{\mathrm s}(e)}$, whose dynamics is represented as
\begin{equation} \label{Dini-W}
D^+ W(t) \leq -\frac{1}{2} b_0 W + \frac{1}{2} \bar \Sigma (\mu, \tau) ,~~~\forall t \geq t_0 + {\mathsf t}(\tau),\end{equation} for $(e, \eta) \in \Omega_{\mathrm T} \times \Sigma$, where $D^+$ is the Dini's upper right derivative. Since $\lim_{t \rightarrow \infty} W(t)\ = \bar \Sigma(\mu, \tau)/b_0$ can be shown in (\ref{Dini-W}), the following should be satisfied:
\begin{equation} \label{cU}
    b_0 \sqrt{c_{\mathrm U}} > \bar \Sigma(\mu, \tau).
\end{equation} It is clear that (\ref{cU}) implies (\ref{sigma bound}) due to $c_{\mathrm T} > c_{\mathrm U}$.

\textit{Lemma 2:} The function $\bar \Sigma(\mu, \tau):{\mathbb R}_{\geq0}\times{\mathbb R}_{>0}\rightarrow {\mathbb R}_{>0}$ has following properties:
\begin{enumerate}
    \item [(a)] For $\mu>0$, $\bar \Sigma(\mu, \tau)$ is strongly convex function of $\tau$. Meanwhile, $\bar \Sigma(0, \tau)$ is a class-$\mathcal{K}$ function of $\tau$.    
    \item [(b)] $\bar \Sigma(\mu, \tau)$ has a global minimum for every $\mu>0$  at $\tau = \{\nu b_1 \sigma_3 \mu / (b_1 \sigma_5 + b_2) \}^{1/(\nu+1)}=: \tau^\dagger(\mu)$. Also, $h(\mu):= \bar \Sigma(\mu, \tau^\dagger(\mu))$ is strictly increasing continuous function of $\mu$, satisfying $\lim_{\mu \rightarrow 0}h(\mu)=0$ and $\lim_{\mu \rightarrow \infty}h(\mu)=\infty$.
    \item [(c)] For arbitrary $k_0>0$, if $\mu\in (0, \mu^\ast)$ where $\mu^\ast:=h^{-1}(k_0)$, then there exist $\underline{\tau}=\underline{\tau}(\mu)<\tau^\dagger(\mu)$ and  $\bar{\tau}=\bar{\tau}(\mu)>\tau^\dagger(\mu)$ such that
    \begin{equation} \label{lem2-2}
         \bar \Sigma(\mu, \tau) < k_0 ~~\Leftrightarrow ~~\tau \in (\underline{\tau}, \bar{\tau}).
    \end{equation}
    \item [(d)] With arbitrary $k_0>0$ and $\mu^\ast = h^{-1}(k_0)$, for $\mu < \mu^\ast$ and $\tau < \tau^\dagger(\mu^\ast)$ if \begin{equation} \label{lem2-3}
        \mu<\frac{(k_0-b_2 \tau-b_1\sigma_5 \tau)\tau^\nu}{b_1\sigma_3 +(b_3 + b_1\sigma_4) \tau^\nu} =: \tilde \mu(k_0, \tau)
    \end{equation} holds, then $\tau$ is larger than $\underline{\tau}(\mu). $
\end{enumerate}

\textit{Proof:} We take the proof to Appendix B. $\square$

Apply (c) of Lemma 2 with $k_0 = b_0 \sqrt{c_{\mathrm U}}$ to have $\mu_1^\ast$ enabling to confirm the existence of $\underline{\tau}_3$ and $\bar \tau_3$ such that (\ref{cU}) is satisfied for all $\tau \in (\underline{\tau}_3, \bar \tau_3)$. Thus, $\dot{V}_{\mathrm s}<0$ for $(e, \eta)\in \partial\Omega_{\mathrm T}\times\Sigma$ by choosing $\tau \in (\underline{\tau}_3, \bar \tau_3)$ if $\mu <\mu^\ast_1$, and it was shown that $\dot{V}_{\mathrm f} < 0 $ for $(e, \eta) \in \Omega_{\mathrm T} \times \partial \Sigma$ for $\bar \tau_1$ in (\ref{fast_invariant}). Hence, the set $\Omega_{\mathrm T} \times \Sigma$ is positively invariant, and $ \lim_{t \rightarrow \infty} V_{\mathrm s} < c_{\mathrm U}$ also holds with $\underline{\tau}_3$, $\bar \tau_3$, and $\mu_1^\ast$.

Therefore, the desired performances are guaranteed with $\underline{\tau} =\underline{\tau}_3$ and $\bar \tau = \min(\bar \tau_1, \bar \tau_2, \bar \tau_3)$ for $\mu<\mu^\ast_1$. However, if either $\varepsilon_{\mathrm T}$ or $\varepsilon_{\mathrm U}$ is much tight, then it may not exist conditioned $\tau$ for some $\mu$. Thus, $\mu^\ast_2:=\tilde \mu(\sqrt{c_{\mathrm U}}/\lambda_{\max}^{\mathrm s}, \min(\bar \tau_1, \bar \tau_2))$ in (\ref{lem2-3}) of Lemma 2 is provided to put lower and upper bounds on $\tau$ in order.
Thus, the proof is completed. ~~~$\blacksquare$

\subsection{Explicit parameter design guideline \\ under measurement noise}

To ensure desired performances (\ref{prob1})--(\ref{prob2}) expressed with $\varepsilon_{\mathrm U}>0$ and $\varepsilon_{\mathrm T}>\varepsilon_{\mathrm U}$, we provide a precise design proposal of $\tau$ depending on noise level $\mu$, which is smaller than specified threshold of noise level $\mu^\ast$, relying on Theorem 1:
 \begin{align*}
     \mu^\ast & = \begin{cases}
     \mu^\ast_1,& \text{if } \min(\bar \tau_1, \bar \tau_2) \geq  \tau^\dagger(\mu_1^\ast), \\
    \mu^\ast_2,              & \text{if } \min(\bar \tau_1, \bar \tau_2) < \tau^\dagger(\mu_1^\ast), \\
\end{cases} \\
\underline{\tau} & =\underline{\tau}_3(\mu), \qquad \qquad  \bar \tau  = \min(\bar \tau_1, \bar \tau_2, \bar \tau_3(\mu)), 
\end{align*} where \begin{subequations} \begin{align}
    \mu_1^\ast & = h^{-1}(\sqrt{c_{\mathrm U}}/\lambda_{\max}^{\mathrm s} ), \label{mu3}\\
    \mu^\ast_2 & =\tilde \mu(\sqrt{c_{\mathrm U}} / \lambda_{\max}^{\mathrm s} , \min(\bar \tau_1, \bar \tau_2)), \label{mu4}\\
    \bar \tau_1 & \; \text{s.t.} \; \forall \tau< \bar \tau_1, \; 4 \tau \lambda_{\max}^{\mathrm f}\sqrt{\check g^2 +1 } \| \check{\mathbf C}_\tau\| <1, \label{tau1}\\
    \bar \tau_2 & \; \text{s.t.} \; \forall \tau< \bar \tau_2, \; {\mathsf t}(\tau) < {\mathsf t}^\ast, \label{tau3}\\
    \underline{\tau}_3 & \;\text{and}\;\bar \tau_3 \; \text{s.t.} \; \forall  \tau \in (\underline{\tau}_3, \bar \tau_3), \;  \sqrt{c_{\mathrm U}} / \lambda_{\max}^{\mathrm s} > \bar \Sigma(\mu, \tau), \label{tau4} 
\end{align} \end{subequations}
for some $c_{\mathrm U}<\lambda_{\min}^{\mathrm s}\varepsilon_{\mathrm U}^2$, and the remaining notations follow the definitions given in Lemma 2 and the proof of Theorem 1. 

As a consequence of $\tau$ design proposal, the following corollary arises: reducing $\tau$ extremely cannot guarantee arbitrary performance except noise-free case.

\textit{Corollary 1:} Under the same assumptions in Theorem 1, for any $\varepsilon_{\mathrm U}>0$ and $\varepsilon_{\mathrm T}>\varepsilon_{\mathrm U}$, a lower bound $\underline{\tau}=\underline{\tau}_3(\mu)$ can be zero only if noise is identically zero, i.e., $\mu = 0$.

\textit{Proof: } As in the proof of Lemma 2, for all $\mu \neq 0$, $\lim_{\tau\rightarrow 0}\bar \Sigma(\mu, \tau)=\infty$ is shown. Then in (\ref{tau4}), $\underline{\tau}_3\neq 0$ for any $c_{\mathrm U}$ and for $\mu \neq 0$. However, $\underline{\tau}_3=0$ can be applied for $\mu=0$, since $\bar \Sigma(0, \tau)$ is a class-$\mathcal{K}$ function of $\tau$ as shown in Lemma 2 (a).

\textit{Remark 2:} In fact, there are not concrete values of $\lambda_{\max}^{\mathrm s}$, $\lambda_{\min}^{\mathrm s}$, $\lambda_{\max}^{\mathrm f}$, and $\lambda_{\min}^{\mathrm f}$, which are maximal and minimal eigenvalues of $P_{\mathrm s}$ and $P_{\mathrm f}$, respectively, due to the uncertainties $S$ and $g$ included in ${\mathbf A}_{\mathrm s}$ and ${\mathbf A}_{\mathrm f}$. In order to derive reliable estimates, we first recommend to run numerous simulations with possible parameters as a realistic approach. Or, if we can believe that ${\mathbf A}_{\mathrm f} + {\mathbf A}_{\mathrm f}^\top$ is also Hurwitz matrix despite of uncertainties, the result of \cite{yasuda1979upper} is able to be applied to confirm a closed interval for eigenvalues of $P_{\mathrm f}$. Similarly, if ${\mathbf A}_{\mathrm s} + {\mathbf A}_{\mathrm s}^\top$ is Hurwitz, the same method can be used for $P_{\mathrm s}$. 

\section{Concluding Remarks}

In this paper, we propose a methodology to determine the parameter inside DOB in order for guaranteeing any required transient and steady state performance. Compared to literature studies to handle the noise sensitivity of DOB, this work has an advantage in terms of compatibility and flexibility by adjusting the single parameter of existing DOB. Also, required noise specification is only its bound, and it does not have to be continuous, differentiable, or dominant in high frequency. As a result, an affordable noise level, which ensures the existence of $\tau$ enabling to achieve desired performance, and novel $\tau$ design guideline are presented. 
Moreover, it is theoretically shown that excessively reducing $\tau$ guarantees to improve performance only under noise-free conditions.

\section*{Appendix}

\subsection*{A. Demonstration on the transformed system (\ref{noisy-singular}) derived by the coordinate change (\ref{coordi})}

For the sake of having $\dot{\xi}$ and $\dot{\zeta}$, we derive $\dot{q} - [\dot x;~0_{l-\nu}]$ by $\dot q$ in (\ref{ss-q}) and $\dot x$ in (\ref{Real plant}) as

\begin{equation*}
\begin{split}
    \dot{q} - [\dot x;0_{l-\nu}] & = A_l\left(q-[x;0_{l-\nu}] \right) \\
    & \quad\,- \frac{a_0}{\tau^l}T_\tau \hat{\mathbf e}_l (\hat {\mathbf e}^\top q-\hat {\mathbf e}^\top x -v) \\
     & \quad\, -\hat{\mathbf e}_\nu \left( \hat{\mathbf e}_\nu^\top\Xi^\ast_\xi-\bar g D v +g{\mathbf d}-g \hat{d} \right) \\
     & = A_l\left(q-[x;0_{l-\nu}] -\hat{\mathbf e}_{\nu+1} \hat{\mathbf e}_\nu^\top\Xi^\ast_\xi \right) \\
      & \quad\, - \frac{a_0}{\tau^l}T_\tau \hat{\mathbf e}_l (q_1-y-v) \\
       & \quad\, -\hat{\mathbf e}_\nu \left(g{\mathbf d}-g \hat{d}  -\bar g D v \right),
\end{split}
\end{equation*}
using $\hat{\mathbf e}_\nu^\top \Xi^\ast_\xi  + g {\mathbf d} = \phi^\top x + \psi^\top z +f_d(x,z) +g u_{\mathrm r}+gd +\bar g D v$, since $\Xi^\ast_\xi$ implies $\hat{\mathbf e}_\nu^\top \Xi_\xi^\ast = \bar{\phi}^\top x + \bar{\psi}^\top \bar{z} + \bar{g} ( u_{\mathrm r}+Dv)$. Thus, $\dot{\xi}$ is organized as follow \begin{equation*} 
\begin{split}
    \tau \dot{\xi} & = \frac{1}{\tau^{\nu}}\Delta \left( \dot{q} - [\dot x;~0_{l-\nu}] \right)-\tau \hat{\mathbf e}_{\nu+1} \hat{\mathbf e}_\nu^\top \dot \Xi^\ast_\xi \\
    & = A_l \xi -a_0T_1\hat{\mathbf e}_l \xi_1 +\frac{a_0}{\tau^\nu} T_1\hat{\mathbf e}_l v \\
     & \quad\, -\hat{\mathbf e}_\nu \left(g{\mathbf d}-g\hat{d} -\bar g D v\right) - \tau \hat{\mathbf e}_{\nu+1} \hat{\mathbf e}_\nu^\top \dot \Xi^\ast_\xi.
    \end{split}
\end{equation*}
Also, $\dot{\zeta}$ is written as 
\begin{equation*}
\begin{split}
    \tau \dot{\zeta} & = \frac{1}{\tau^l} \Delta\dot{p}-\tau \hat{\mathbf e}_1 \dot{\Xi}^\ast_\zeta, \\
     & = \frac{1}{\tau^l} \Delta \left( A_lp+\hat{\mathbf e}_l \left( - \frac{a_0}{\tau^l} \hat{\mathbf e}_1^\top T_\tau p + u_{\mathrm r} -\hat{d} \right)\right) -\tau \hat{\mathbf e}_1  \dot{\Xi}^\ast_\zeta \\
     & = A_l\zeta+\hat{\mathbf e}_l\left( - a_0 \hat{\mathbf e}_1^\top T_1 \zeta-a_0 \Xi^\ast_\zeta + u_{\mathrm r} -\hat{d} \right)-\tau \hat{\mathbf e}_1 \dot{\Xi}^\ast_\zeta \\
    & = A_l\zeta-a_0\hat{\mathbf e}_l\hat{\mathbf e}_1^\top T_1 \zeta+\hat{\mathbf e}_l\left({\mathbf d} -\hat{d} -\frac{\bar g}{g} D v\right)- \tau \hat{\mathbf e}_1 \dot{\Xi}^\ast_\zeta.
    \end{split}
\end{equation*}

Therefore, the overall closed-loop dynamics partitioned with $\xi$, $\zeta$ and $e$ is represented as
\begin{subequations} \label{noisy-semi-singular}
\begin{align}
    \tau \dot{\xi} & = A_l \xi -a_0T_1\hat{\mathbf e}_l \xi_1 -\hat{\mathbf e}_\nu (g{\mathbf d}-g\hat{d} -\bar g D v) \label{xi-dyna1} \\
     & \quad\, +\frac{a_0}{\tau^\nu} T_1\hat{\mathbf e}_l v - \tau \hat{\mathbf e}_{\nu+1} \hat{\mathbf e}_\nu^\top \dot \Xi^\ast_\xi, \nonumber \\
    \tau \dot{\zeta} & = A_l\zeta-a_0\hat{\mathbf e}_l\hat{\mathbf e}_1^\top T_1 \zeta+\hat{\mathbf e}_l\left({\mathbf d} -\hat{d} -\frac{\bar g}{g} D v\right)- \tau \hat{\mathbf e}_1 \dot{\Xi}^\ast_\zeta,  \label{zeta-dyna1}\\
    \dot{e} & = {\mathbf A}_{\mathrm s}e +g \hat{\mathbf e}_\nu \left( {\mathbf d}- \hat{d}\right) - N v. \label{error1}
\end{align} \end{subequations}

As long as $|w-y_p| \leq \bar s$, namely, the saturation function $\Pi$ is inactive, $\hat{d} = w-y_p$ holds. Thus, $w$ and $y_p$ in (\ref{w}) and (\ref{yp}) are needed to be rewritten with the transformed states $\xi$ and $\zeta$ by utilizing the following notes.

\textit{Note 1:} For a matrix $\Gamma_\tau $ of the following structure
\begin{equation*}
    \Gamma_\tau = \begin{bmatrix}
        \gamma_1 & \gamma_2 \tau & \cdots &  \gamma_{j-1} \tau^{j-2} & \gamma_j \tau^{j-1} \\
        0 & \gamma_1 & \cdots & \gamma_{j-2} \tau^{j-3} & \gamma_{j-1} \tau^{j-2} \\
        \vdots & \ddots & \ddots & \ddots & \vdots \\
        0 & \cdots  & 0 & \cdots & \gamma_{j-i+1} \tau^{j-i}  \\
    \end{bmatrix}~~\in {\mathbb R}^{i \times j},
\end{equation*} where $0<i\leq j$ and $\{\gamma_1,\gamma_2,...,\gamma_j \}\subset {\mathbb R}$, the below equality is hold:
\begin{equation*}
    \Gamma_\tau \diag(\tau^{-1},...,\tau^{-j}) = \diag(\tau^{-1},...,\tau^{-i})\Gamma_1.
\end{equation*}

\textit{Note 2:} If $\Gamma_\tau$ defined in Note 1 is invertible, then $\Gamma^{-1}_\tau$ also has the same structure of $\Gamma_\tau$ with respect to $\tau$.

We share the fact that $C_\tau$, $\bar C_\tau$, and $M_l^\nu (C_\tau)$ have the matrix structure with respect to $\tau$ introduced in Note 1.

First, $y_p$ in (\ref{yp}) applied the transformation (\ref{coordi2}) becomes \begin{equation*}
\begin{split}
    y_p & = \frac{a_0}{\tau^l}C_\tau \left( \tau^{l+1} \diag(\tau^{-1},...,\tau^{-l}) \zeta + \tau^l \hat{\mathbf e}_1\Xi^\ast_\zeta \right) \\
    & = a_0 C_1 \zeta + a_0 \Xi^\ast_\zeta,
    \end{split}
\end{equation*} with $C_\tau$ structure with respect to $\tau$ and Note 1.

Next, for the sake of constructing $w$ in (\ref{w}) with $\xi$, the terms inside $w$ associated with $q$ are organized \begin{align} 
    M_l^\nu(C_\tau) T_\tau ^{-1}q & = \tau^{\nu+1} M_l^\nu(C_\tau) T_\tau ^{-1} \diag(\tau^{-1},...,\tau^{-l}) \xi \nonumber \\
    & \quad \,+ M_l^\nu(C_\tau) T_\tau ^{-1} \left( \hat{\mathbf e}_{\nu+1} \hat{\mathbf e}^\top_\nu \Xi^\ast_\zeta + [x; \; 0_{l-\nu}] \right) \nonumber \\ 
    &= \diag(\tau^\nu,...,\tau) M_l^\nu(C_1)T_1 ^{-1} \xi \nonumber \\
    & \quad \, + M_l^\nu(C_\tau)T_\tau ^{-1} \left(\hat{\mathbf e}_{\nu+1}\hat{\mathbf e}_{\nu}^\top \Xi^\ast_\xi  + [x; ~ 0_{l-\nu}] \right), \label{w-1-1} \\
        C_\tau A_l^\nu T_\tau ^{-1} q & = C_1 A_l^\nu T_1 ^{-1}\xi \nonumber \\
        & \quad \, +C_\tau A_l^\nu T_\tau ^{-1} \left(\hat{\mathbf e}_{\nu+1}\hat{\mathbf e}_{\nu}^\top \Xi^\ast_\xi + [x; ~ 0_{l-\nu}] \right) \nonumber \\
        & = C_1 A_l^\nu T_1 ^{-1}\xi+ \hat{\mathbf e}_{\nu}^\top \Xi^\ast_\xi, \label{w-1-2}
\end{align} utilizing $C_\tau A_l^\nu T_\tau ^{-1} [x; ~ 0_{l-\nu}]  = 0$ and Note 1 in both (\ref{w-1-1}) and (\ref{w-1-2}). For (\ref{w-1-1}). Let us recall $\bar C_\tau = C_\tau - Z^\top_1 T_\tau$, then the following arises \begin{equation*}
    M_l^\nu(C_\tau)T_\tau ^{-1} = M_l^\nu(\bar C_\tau)T_\tau ^{-1}+\begin{bmatrix}
        I_\nu ~~0_{\nu \times(l-\nu)}
    \end{bmatrix}. 
\end{equation*} Then, the following is obtained by premultiplying the row vector $\bar \phi^\top$ to (\ref{w-1-1}) \begin{multline}
    \bar{\phi}^\top M_l^\nu(C_\tau)T_\tau ^{-1}q = \tau \bar g {\mathbf C}_\tau \xi + \bar \phi^\top x \\ 
    + \tau \bar{\phi}^\top M_l^\nu(\tau^{-1} \bar C_\tau)T_\tau ^{-1} \left(\hat{\mathbf e}_{\nu+1}\hat{\mathbf e}_{\nu}^\top \Xi^\ast_\xi  + [x; ~ 0_{l-\nu}] \right), \label{w-deri1}
\end{multline}  with $\tau M_l^\nu(\tau^{-1} \bar C_\tau) = M_l^\nu(\bar C_\tau)$.

Depending on the structure of Byrnes-Isidori normal form \cite{byrnes1991asymptotic}, the following holds \begin{multline}
    \hat{\mathbf e}_{\nu+1}\hat{\mathbf e}_{\nu}^\top \Xi^\ast_\xi  + [x; ~0_{l-\nu}] \\ = [\hat{\mathbf e}_2 \,\cdots \, \hat{\mathbf e}_{\nu+1} \, 0_{l \times (n_{\mathrm e} -\nu)}] \Xi^\ast_\xi
    + [x_1;~ 0_{l-1}], \label{w-deri2}
\end{multline} invoking $\Xi^\ast_\xi  ={\mathbf A}_{\mathrm s} e-\dot \chi_{\mathrm n}$. Since the first entry of the row vector $\bar C_\tau$ is zero and $T_\tau^{-1}$ is upper triangular, $\bar C_\tau T_\tau^{-1} [x_1;~ 0_{l-1}] = 0$ holds. Then, we have $w$ in terms of $\xi$ by (\ref{w-1-2})--(\ref{w-deri2}) as follows \begin{equation*}
\begin{split}
    w & = \frac{1}{\bar{g}}\left( -\bar{\psi}^\top \bar{z}- \bar \phi^\top x+ \hat{\mathbf e}_{\nu}^\top \Xi^\ast_\xi \right) \\
    & \quad \, -\frac{1}{\bar{g}}\left(\tau\bar g {\mathbf C}_\tau  \xi  +\tau\bar g \bar{\mathbf C}_\tau  \Xi^\ast_\xi - C_1 A_l^\nu T_1 ^{-1} \xi \right) \\
    & = u_{\mathrm r} -\tau {\mathbf C}_\tau  \xi  - \tau\bar{\mathbf C}_\tau  \Xi^\ast_\xi + \frac{1}{\bar{g}}C_1 A_l^\nu T_1 ^{-1} \xi + D v.
    \end{split} 
\end{equation*} The first equality also relies on the definition of $\bar{\mathbf C}_\tau$, and the second equality is from $\hat{\mathbf e}_{\nu}^\top \Xi_\xi^\ast = \bar{\phi}^\top x + \bar{\psi}^\top \bar{z}  +\bar{g} \left( u_{\mathrm r}+D v\right)$. Hence, $w-y_p$ is obtained as \begin{equation}
\begin{split} \label{d esti.}
    w-y_p & = \left( u_{\mathrm r} -\tau {\mathbf C}_\tau  \xi  - \tau\bar{\mathbf C}_\tau  \Xi^\ast_\xi + \frac{1}{\bar{g}}C_1 A_l^\nu T_1 ^{-1} \xi + D v \right) \\
    & \quad \, - \left(a_0 C_1 \zeta + a_0 \Xi^\ast_\zeta \right) \\ 
    & = {\mathbf d}+\frac{1}{\bar{g}} C_1 A_l^\nu T_1^{-1} \xi - a_0C_1\zeta +\frac{g-\bar{g}}{g}D v \\
    & \quad \,  - \tau \left( {\mathbf C}_\tau \xi +  \bar{\mathbf C}_\tau \Xi^\ast_\xi\right),
\end{split}
\end{equation} by $\Xi^\ast_\zeta = (u_{\mathrm r}- {\mathbf d} + (\bar g D v)/g)/a_0$. As a result, manipulate the equations (\ref{noisy-semi-singular}) for $\dot \xi$, $\dot \zeta$, and $\dot e$ by $\hat d = w-y_p$ in (\ref{d esti.}):
\begin{align*}
    \tau \dot{\xi} & = A_l \xi -a_0T_1\hat{\mathbf e}_l \xi_1 -\hat{\mathbf e}_\nu (g{\mathbf d}-g(w-y_p)) -\bar g D v) \\
    & \quad \, +\frac{a_0}{\tau^\nu} T_1\hat{\mathbf e}_l v - \tau \hat{\mathbf e}_{\nu+1} \hat{\mathbf e}_\nu^\top \dot \Xi^\ast_\xi \\
    &  =  {\mathbf A}_{\mathrm 11} \xi + {\mathbf A}_{\mathrm 12} \zeta  -\tau g\hat{\mathbf e}_\nu \left( {\mathbf C}_\tau \xi  + \bar{\mathbf C}_\tau \Xi_\xi^\ast \right) \\
    & \quad \, - \tau \hat{\mathbf e}_{\nu+1} \hat{\mathbf e}_\nu^\top \dot \Xi^\ast_\xi + \frac{a_0}{\tau^\nu}T_1 \hat{\mathbf e}_l v  + {\mathbf N}_\xi v, 
\end{align*}
\begin{align*}
    \tau \dot{\zeta} & = A_l\zeta-a_0\hat{\mathbf e}_l\hat{\mathbf e}_1^\top T_1 \zeta \\
    & \quad \, +\hat{\mathbf e}_l\left({\mathbf d} -(w-y_p) -\frac{\bar g}{g} D v\right)- \tau \hat{\mathbf e}_1 \dot{\Xi}^\ast_\zeta \\
    & = A_l\zeta-a_0\hat{\mathbf e}_l\hat{\mathbf e}_1^\top T_1 \zeta - \tau \hat{\mathbf e}_1 \dot{\Xi}^\ast_\zeta - D\hat{\mathbf e}_l v \\
    & \quad \, -\hat{\mathbf e}_l\left(\frac{1}{\bar{g}} C_1 A_l^\nu T_1^{-1} \xi - a_0C_1\zeta  - \tau \left( {\mathbf C}_\tau \xi + \bar{\mathbf C}_\tau \Xi^\ast_\xi\right) \right) \\
     & = {\mathbf A}_{\mathrm 21} \xi + {\mathbf A}_{\mathrm 22}\zeta +\tau \hat{\mathbf e}_l\left({\mathbf C}_\tau \xi  + \bar{\mathbf C}_\tau \Xi_\xi^\ast\right) - \tau \hat{\mathbf e}_1 \dot{\Xi}^\ast_\zeta + {\mathbf N}_\zeta v,
\end{align*} and
\begin{align*}
    \dot{e} & = {\mathbf A}_{\mathrm s}e +g \hat{\mathbf e}_\nu \left( {\mathbf d}- (w-y_p) \right) - N v \\
    & = {\mathbf A}_{\mathrm s}e - g \hat{\mathbf e}_\nu \left(\frac{1}{\bar{g}} C_1 A_l^\nu T_1^{-1} \xi - a_0C_1\zeta  - \tau \left( {\mathbf C}_\tau \xi +  \bar{\mathbf C}_\tau \Xi^\ast_\xi\right) \right)\\
    & \quad \, +(\bar g - g) D \hat{\mathbf e}_\nu v -Nv\\
    & = {\mathbf A}_{\mathrm s}e + {\mathbf F}_{\mathrm 1}\xi + {\mathbf F}_{\mathrm 2}\zeta + \tau g\hat{\mathbf e}_\nu \left( {\mathbf C}_\tau \xi  + \bar{\mathbf C}_\tau \Xi_\xi^\ast \right) + {\mathbf N}_{\mathrm e}v. \label{noisy-perfor}
\end{align*} Thus, the claim is proved.

Also for the proof of Lemma 1, let $v \equiv 0$ be in the transformed system (\ref{noisy-singular}), then the standard perturbation form shown in Lemma 1 can be obtained.

\subsection*{B. Proof for Lemma 2}

Calculate the first and second partial derivatives of $\bar \Sigma(\mu, \tau)$ with respect to $\tau$, then we have \begin{equation} \label{1st/2nd}
    \pdv{\bar \Sigma}{\tau} = b_2   -\frac{\nu b_1 \sigma_3 }{\tau^{\nu+1}}\mu + b_1\sigma_5,~ \pdv[2]{\bar \Sigma}{\tau} =\frac{\nu (\nu+1)b_1 \sigma_3 }{\tau^{\nu+2}}\mu.
\end{equation} It is shown that for $\mu>0$, $\bar \Sigma(\mu, \tau)$ is strongly convex function of $\tau$ due to $\partial^2\bar \Sigma / \partial \tau^2 > 0$ for all $\tau>0$. Also, $\bar \Sigma(0, 0)$ holds by definition, and $\bar \Sigma(0, \tau)$ is strictly increasing function of $\tau$ due to $\partial\bar \Sigma / \partial \tau > 0$ for all $\tau>0$. Hence (a) is clear.

Further, (b) is proved due to $\partial\bar \Sigma / \partial \tau = 0$ on $\tau = \tau^\dagger(\mu) = \{\nu b_1 \sigma_3  \mu /  (b_2 +b_1 \sigma_5) \}^{1/(\nu+1)}$ in (\ref{1st/2nd}). By substituting $\tau^\dagger(\mu)$ into $\bar \Sigma$, it is clear that $h(\mu)$ is radially unbounded strictly increasing function and has $\lim_{\mu \rightarrow 0}h(\mu) = 0$, i.e., $h^{-1}: {\mathbb R}_{>0} \rightarrow {\mathbb R}_{>0}$ is bijection.

Next, existence of $\mu^\ast=h^{-1}(k_0)$ for arbitrary given $k_0>0$ is relying on bijection $h^{-1}$. Since $\bar \Sigma(\mu, \tau)$ has both $\tau^{-\nu}$ and $\tau$ order terms for all $\mu \neq 0$, it is clear that $\lim_{\tau \rightarrow 0}\bar \Sigma(\mu, \tau)=\infty$, and $\lim_{\tau \rightarrow \infty}\bar \Sigma(\mu, \tau)=\infty$. Hence, for $\mu\in (0, \mu^\ast)$, there are unique $\underline{\tau}(\mu)\neq 0$ and $\bar \tau(\mu)>\tau^\dagger (\mu)$ so that (\ref{lem2-2}) is satisfied, depending on the continuity of $\bar \Sigma$, (a), and (b).

At last, (d) is easily obtained, after reorganizing the inequality $\bar \Sigma(\mu, \tau) < k_0$.

%

\bibliographystyle{ieeetr}
\bibliography{gaeunbib}


\end{document}